\documentclass[pre,aps,twocolumn,superscriptaddress,floatfix]{revtex4}

\usepackage{amsmath}
\usepackage{amssymb}
\usepackage{amsbsy}
\usepackage{graphicx}
\usepackage{color}

\usepackage{enumerate}
\usepackage{mathtools}

\begin{document}

\def\be{\begin{equation}}
\def\ee{\end{equation}}     
\def\bfi{\begin{figure}}
\def\efi{\end{figure}}
\def\bea{\begin{eqnarray}}
\def\eea{\end{eqnarray}}
\def\darkgreen{\textcolor[rgb]{0,.3,0}}

\newcommand{\ra}{\rightarrow}
\newcommand{\scatt}{{\rm scatt}}
\newcommand{\prob}{P}
\newcommand{\tot}{{\rm tot}}
\newcommand{\stat}{{\rm stat}}
\newcommand{\erfc}{{\rm erfc}}
\newcommand{\mE}{\mathcal{E}}

\title{Probability distributions with singularities}

\author{Federico Corberi}
\address{Dipartimento di Fisica ``E.~R. Caianiello'', Universit\`a  di Salerno, 
via Giovanni Paolo II 132, 84084 Fisciano (SA), Italy.}
\address{INFN, 
Gruppo Collegato di Salerno, and CNISM, Unit\`a di Salerno, Universit\`a  di Salerno, 
via Giovanni Paolo II 132, 84084 Fisciano (SA), Italy.}

\author{Alessandro Sarracino}
\address{Dipartimento di Ingegneria, Universit\`a della Campania ``L. Vanvitelli'', 81031 Aversa (CE), Italy}

\begin{abstract}
  In this paper we review some general properties of probability distributions which exibit a singular behavior.
After introducing the matter with several examples based on various models of statistical mechanics,  we 
discuss, with the help of such paradigms, the underlying mathematical mechanism producing the singularity
and other topics such as the condensation of fluctuations, 
the relationships with ordinary phase-transitions, the giant response associated to anomalous
fluctuations, and the interplay with Fluctuation Relations.
\end{abstract}

\maketitle

\section{Introduction} \label{intro}

Quantitative predictions on the occurrence of rare events can be very
useful particularly when these events can produce macroscopic effects
on the system.  This occurs, for instance, when a large fluctuation
triggers the decay of a metastable state~\cite{langer} leading the
system to a completely different thermodynamic condition. Other
examples with rare deviations producing important effects are found in
many other contexts, as in information theory~\cite{kafri} and
finance~\cite{filiasi}.

For a collective variable $N$, namely a quantity formed by the
addition of many microscopic contributions, such as the energy of a
perfect gas or the mass of an aggregate, typical fluctuations are
regulated by the Central Limit Theorem.  Rare events, instead, may go
beyond the theorem's validity and are described by large deviations
theory \cite{touchette2009,vulp} which, in principle, aims at
describing the whole spectrum of possible fluctuations, no matter how
large or rare they are, by means of their full probability
distribution $P(N)$.

It has been found that, in many cases, $P(N)$ exhibits a singular
behavior, in that it is non-differentiable around some value (or
values) $N_c$ of the fluctuating
variable~\cite{filiasi,magnetic1,magnetic2,magnetic3,magnetic4,memory,gradenigo2013,qquench,disordered,noldp1,noldp2,noldp3,noldp4,noldp5,noldp6,noldp7,noldp8,noldp9,noldp10,noldp11,oth1,oth2,oth3,oth4,oth5,oth6,oth7,oth8,gaussian1,gaussian2,gaussian3,heatexch,crisanti,zan,cagnetta2017}.
Such singularities have an origin akin to those observed in the
thermodynamic potentials of systems at
criticality. Indeed, a correspondence can be shown between $P(N)$ and
the free energy of a companion system, related to the one under study
by a duality map~\cite{touchette2009, gaussian1,gaussian2,gaussian3}, which is interested
by a phase-transition.

Recently, a great effort has been devoted to the characterization of
these singular behaviors in the large deviations functions of
different models where analytical results can be obtained. This has
unveiled a rich phenomenology which shares common features. In most
cases non-analycities are a consequence of a particular condensation
phenomenon denoted as {\it condensation of fluctuations}.  It occurs
when a significant contribution to the fluctuations is built within
a limited part of phase-space, or is provided by just one of the
degrees of freedom of the system. This is analogous to what happens,
for instance, in the usual condensation of a gas when it concentrates
in a liquid drop, or in the well-known Bose-Einstein condensation,
where the mode with vanishing wavevector contributes
macroscopically. However, while usual condensation represents the
typical behavior of the system, the condensation of fluctuations can
only be observed when certain rare events take place.

Another interesting feature of systems with singular probability
distributions can be their extreme sensibility to small
perturbations. Usually, the properties of a system made of many
constituents or degrees of freedom do not change much if some features
of a single particle are slightly changed. This is true both for the
average properties and for the fluctuations. For instance, neither the
average energy of a gas nor its fluctuations change appreciably if the
mass of one single molecule is increased a bit. This is simply because
this particle is only one out of an Avogadro number.  However, when
condensation of fluctuations occurs, one can observe a giant response
if the perturbed degree of freedom is exactly the one that contributes
macroscopically to the fluctuation.

Singular probability distributions raise the question about the
validity of the Fluctuation Relations (FRs).  These relations have
been extensively studied recently~\cite{umberto,seifert} because they
reflect general symmetries of the deviations of certain quantities
and are believed to contribute to a general understanding of
non-equilibrium states. In particular, FRs connect the probability of
observing events with a certain value $N$ of the fluctuating variable,
to the probability of the events associated to the opposite value $-N$.  Among
other open issues on the subject, one is represented by the case of
singular fluctuations.  Indeed, the singularity in $N_c$ usually
separates two regions where fluctuations have very different
properties.  For instance, on one side of $N_c$ one can have a
standard situation where all the degrees of freedom contribute,
whereas on the other side fluctuations can condense and be determined
by the contribution of a single degree. Clearly, if $N$ is such that
$N$ and $-N$ fall on different branches of $P(N)$, namely on the two
sides of $N_c$, the mechanism whereby an FR can be fulfilled must be
highly non-trivial.  In general, singular probability distributions
may, or may not, exhibit the FR and a general understanding of this
point is still not achieved.

This paper is a brief review devoted to the discussion of singular
probability distributions where, without any presumption of
completeness neither of mathematical rigor, we present examples of
models where such non-analycities show up, we highlight the
mathematical mechanism producing condensation, and we discuss some
relevant aspects related to the subject, such as those mentioned
above. We do that in a physically oriented spirit, providing whenever
possible an intuitive interpretation and a simple
perspective. Non-differentiable probability distributions have been
previously reviewed also in~\cite{baek}, where however the authors
focus on different models and complementary aspects with respect to
those addressed in this paper.

The paper is organized as follows. In section \ref{generalities} we
recall some basic results of probability theory and introduce some
notations. In section \ref{examples} we present some models of
statistical mechanics where non-differentiable probability
distributions have been computed for different collective
quantities. In section \ref{general} we illustrate in detail some
phenomena related to the singular distribution function, mainly using the urn
model as a paradigm, and discuss how similar behaviors arise in other systems. 
We also discuss the topic of the fluctuation relations. More
specific features, such as giant response and observability, are then
presented in section \ref{peculiar}, and, finally, some conclusions
are drawn in section \ref{conclusions}.

\section{Probability distributions: generalities}
\label{generalities}

We consider a generic stochastic system, whose physical state is
defined by the random variable $x$ taking values on a suitable phase
space.
We will be mainly interested in the behavior of collective random
variables, that are defined as the sum of a large number of
{\it microscopic} random variables. For these quantities some general
results can be derived~\cite{vulp}.
As an example let us consider the sum $N=\sum_{j=1}^Mx_j$ 
of a sequence of $M$ random variables $x_j$,
with empirical mean
\begin{equation}
\rho=\frac{N}{M}=\frac{1}{M}\sum_{j=1}^Mx_j.
\label{defN}
\end{equation}
The quantities $x_j$ can represent a sequence of states of a system (for
instance, the position of a particle along a trajectory) or an ensemble of variables
describing its microscopic constituents (e.g., the energies of the
single particles of a gas).  In the case of independent identically
distributed variables, with expectation $\langle x\rangle$ and finite
variance $\sigma$, one has that the empirical mean tends to $\langle
x\rangle$ for large $M$, namely
\begin{equation}
\lim_{M\to\infty}p(\rho-\langle x\rangle < \epsilon)\to 1,
\label{lln}
\end{equation}
where $\epsilon$ is a small quantity and hereafter $p(E)$ (also $P(E)$
or ${\cal P}(E)$) is the probability of an event $E$. The above
equation represents the Law of Large Numbers.

As a further step, one can describe the statistical behavior of the
small fluctuations of $\rho$ around the average $\langle \rho\rangle$,
$\delta \rho=\rho-\langle \rho\rangle$, introducing the quantity
\begin{equation}
z_M=\frac{1}{\sigma \sqrt{M}}\sum_{j=1}^M(x_j-\langle x\rangle),
\end{equation}
which, for very large $M$, and for $\delta \rho \lesssim O(\sigma/\sqrt{M})$, 
has the following distribution function
\begin{equation}
p(z_M=z)\simeq \frac{1}{\sqrt{2\pi}}e^{-\frac{z^2}{2}}.
\label{clt}
\end{equation}
This results is the Central Limit Theorem (CLT), that holds also in
the case of weakly correlated variables.  

More in general, fluctuations of arbitrary size of the quantity $\rho$
can, under certain conditions, be characterized by the Large Deviation Principle (LDP)
\begin{equation}
p(\rho=y)\sim e^{-MI(y)},
\label{ldp}
\end{equation}
where $I(y)$ is the so called rate function. When $p(\rho)$ has a
single absolute maximum (in $\langle \rho\rangle$), the rate function
is positive everywhere but for $y=\langle \rho\rangle$, where it
vanishes.  It is easy to obtain the CLT (\ref{clt}) from the LDP
(\ref{ldp}) by expanding up to second order the function $I(y)$ around
$\langle \rho\rangle$. However, as we will discuss in detail below,
there are interesting cases where the LDP in the form~(\ref{ldp}) is
not satisfied.

A simple example where LDP holds and the rate function can be easily
computed is obtained by considering $\{x_j\}$ as dichotomous variables
taking the value $+1$ with probability $q$ and $-1$ with probability
$1-q$. Then, using the Stirling approximation, one obtains
the explicit expression for the rate function:
\begin{equation}
I(y)=\frac{1+y}{2}\ln \frac{1+y}{2 q}+
\frac{1-y}{2}\ln \frac{1-y}{2(1- q)}. 
\label{cramer}
\end{equation}
Expanding Eq.~(\ref{cramer}) around the mean $\langle y\rangle=2q-1$ one has the CLT
\begin{equation}
I(y)\simeq \frac{(y-\langle y\rangle)^2}{2 (1-\langle y\rangle)}.
\label{cramer2}
\end{equation}

\section{Singular probability distributions: examples} \label{examples}

As far as small deviations of a collective variable are considered,
the associated probability distribution is usually regular,
being a Gaussian when the hypotheses of the CLT are satisfied.
Moving to the realm of large deviations, instead, can hold surprises as,
for instance, the emergence of non-analycities.
Before deepening the meaning and the bearings of the singular behavior,
in this Section we first itemize some
examples of systems where it has been observed. We will then study
it in more detail in some specific models in the following sections.

\subsection{Gaussian model} \label{thegaussianmodel}

The Gaussian model is a reference model of statistical mechanics.
An order-parameter field $\phi(\vec x)$ (which in the magnetic language can be
thought of as a local magnetization at site $\vec x$) is ruled by the following Hamiltonian
\be
   {\cal H}[\varphi]=\frac{1}{2}\int _V d\vec x\, [(\nabla \varphi)^2+r\varphi^2
     (\vec x)],
   \label{hamgaussian}
\ee
where $r>0$ is a parameter and $V$ the volume.
This simple model can be exactly solved and has a rather trivial
phase-diagram without phase transitions.

\begin{figure}[ht]
\vspace{2cm}
\centering
\rotatebox{0}{\resizebox{.45\textwidth}{!}{\includegraphics{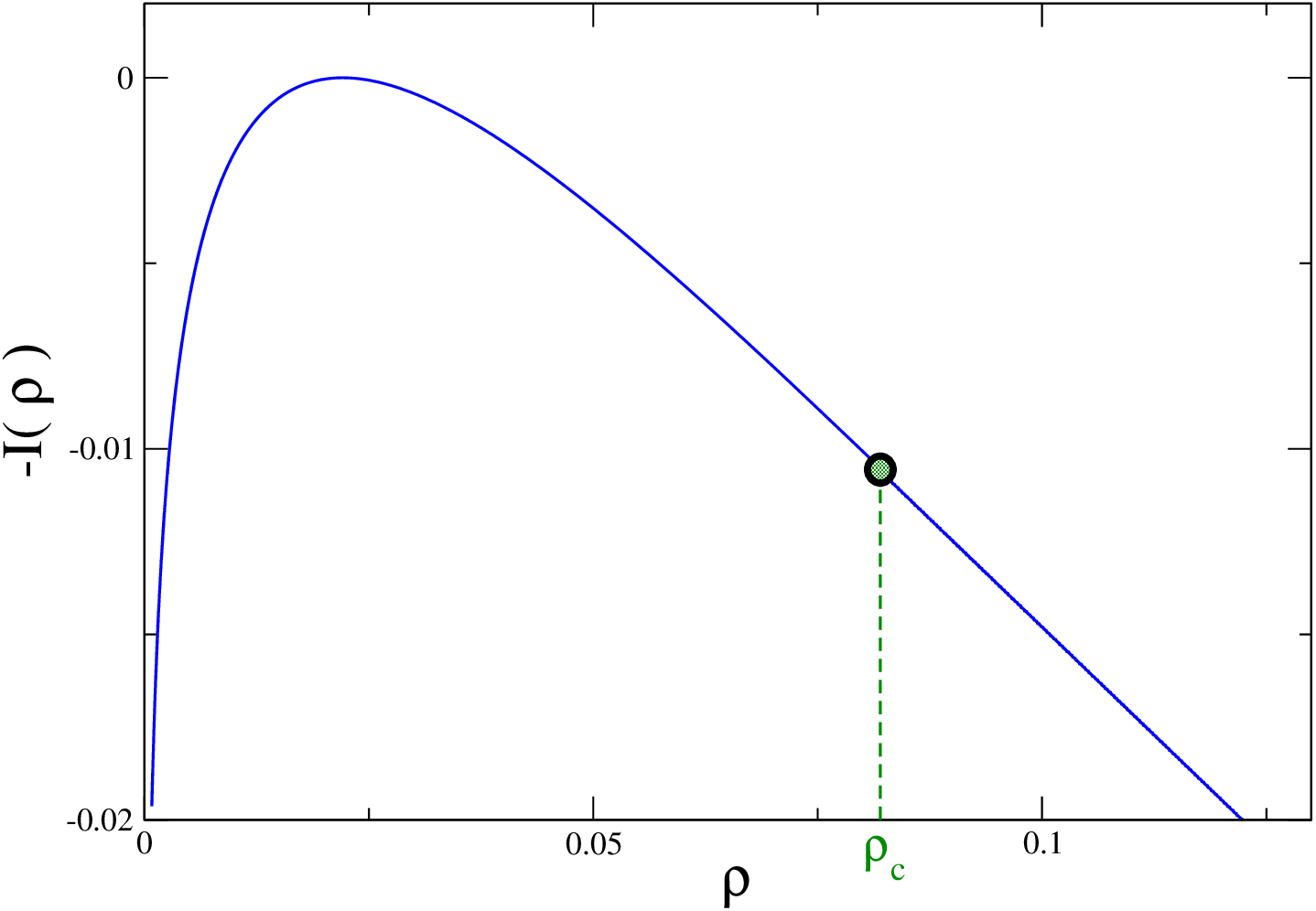}}}
\caption{The (negative) rate function $I(\rho)$ 
  of the variance $N$ of the order parameter field in the Gaussian
  model in $d=3$, with $r=1$, in equilibrium at the temperature $T=0.2$.}  
\label{fig_gauss}
\end{figure}

Let us consider the collective variable
\be
N[\varphi]=\int _V d\vec x \, \varphi ^2 (\vec x),
\label{defvar}
\ee
namely the order parameter variance, and its density $\rho=N/V$.
Its probability distribution was computed analytically in \cite{gaussian1,gaussian2,gaussian3}.
The (negative) rate function of this quantity,
evaluated in equilibrium at a given temperature
$T$, is plotted in Fig.~\ref{fig_gauss}. 
The curve has a maximum in correspondence to the most probable value, where $I(\rho)$ vanishes.
Far from such maximum, in the large deviations regime, the rate function
exhibits a singularity (marked with a green dot) at $\rho =\rho _c$.
In this point the third derivative of the rate function has a discontinuity \cite{gaussian1,gaussian2,gaussian3}.
The existence of such a singularity is related to the fact that, as we will discuss later,
fluctuations with $\rho >\rho_c$ have a different character with the respect to the 
ones in the region $\rho <\rho_c$ where the average, or typical, behavior of the system 
(i.e. the most probable value of $\rho$) is located.

\subsection{Large-${\cal N}$ model}
\label{heatlargen}

Another reference model of statistical mechanics is the description of
a magnetic system in terms of the Ginzburg-Landau Hamiltonian
\be
   {\cal H}[\varphi]=\frac{1}{2}\int _V d\vec x\, \left [
     (\nabla \varphi)^2+r\varphi^2
     (\vec x) +\frac{g}{2{\cal N}} (\varphi ^2)^2\right ],
   \label{hamgl}
\ee
where the ${\cal N}$-components vectorial field $\varphi$ has a
meaning similar to that of the Gaussian model, and $r<0$ and $g>0$ are parameters.
In the large-${\cal N}$ limit (sometimes also denoted as {\it spherical limit})
the model is exactly soluble. There is a phase transition at a
finite critical
temperature $T_c$ separating a paramagnetic phase for $T>T_c$ from a
ferromagnetic one at $T<T_c$.

\begin{figure}[ht]
\vspace{2cm}
\centering
\rotatebox{0}{\resizebox{.45\textwidth}{!}{\includegraphics{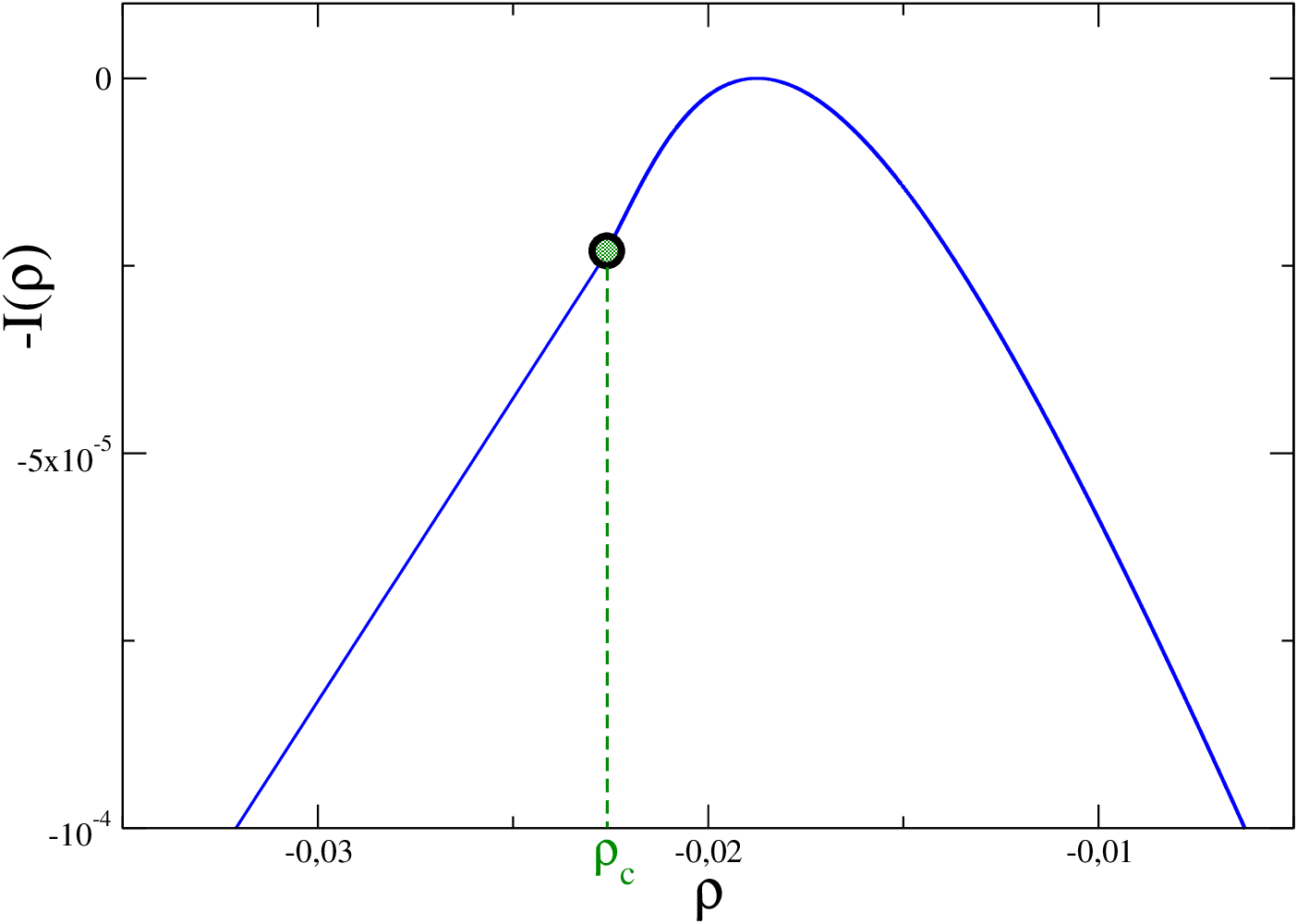}}}
\caption{The (negative) rate function $I(\rho)$ of the probability
  distribution $P(N)$
  of the energy $N$ exchanged by the large-${\cal N}$ model  
  in $d=3$, with $g=-r=1$, with the environment after a quench to zero
  temperature.}  
\label{fig_heatexch}
\end{figure}

The probability distribution of the
energy $N(t,t_w)={\cal H}[\varphi,t]-{\cal H}[\varphi,t_w]$
exchanged by the system in a time interval $[t_w,t]$ with a thermal bath
was computed exactly in \cite{heatexch}.
The (negative) rate function of the intensive quantity $\rho (t,t_w)=N(t,t_w)/V$ is shown in Fig. \ref{fig_heatexch}. This figure refers to the case of a system quenched from a very high temperature to another $T<T_c$. Also in this case there is a singularity 
corresponding to a certain value the quantity $\rho (t,t_w)= \rho_c$ where the third derivative has a discontinuity, and this reflects a different mechanism of heat exchanges for $\rho <\rho_c$ and
for $\rho >\rho_c$.

\subsection{Urn model} \label{urnmodel}

\begin{figure}[ht]
\vspace{2cm}
\centering
\rotatebox{0}{\resizebox{.45\textwidth}{!}{\includegraphics{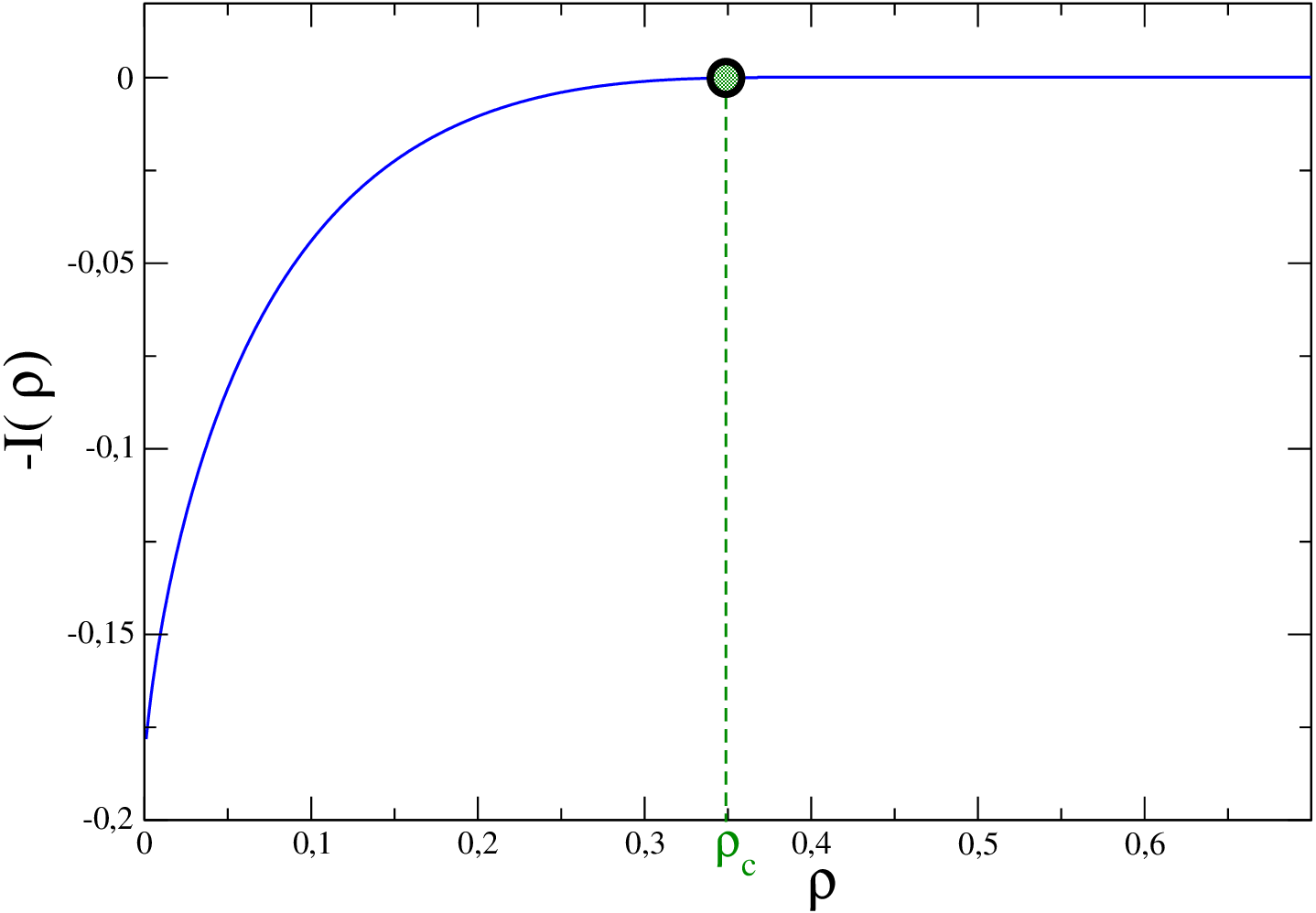}}}
\caption{The rate function $I(\rho)$ of the probability
  distribution $P(N)$
  of the total number of particles $N$ in the urn
  model with $k=3$.} 
\label{fig_urns}
\end{figure}

Let us consider a set of integer
variables $n_i\geq 0$ ($i=1,\dots,M$)
equally distributed in such a way that the probability of having a certain
value $n$ of $n_i$ is
\be
p(n)=\zeta ^{-1}(n+1)^{-k},
\label{urnmicroprob}
\ee
where $\zeta $ is a normalization constant and $k$ a parameter.
One can think of having $M$ urns, each of them
hosts a quantity $n_m$ of particles taken with probability (\ref{urnmicroprob})
from a reservoir. This setting is appropriate to describe a wealth of
situations in many areas of science, from network dynamics to financial data.
The probability distribution of the total number of particles
\be
N=\sum _{m=1}^M n_m
\label{totparticles}
\ee
was studied for large $M$ in different contexts \cite{noldp1,noldp4,noldp8,noldp9,noldp10,corb_dyn}.
The (negative) rate function is shown in Fig. \ref{fig_urns}.
Also in this model it is found that, if $k>2$, there is a singularity at $\rho =\rho _c$, that in this particular case coincides with the average value $\langle \rho\rangle$. Notice that in this case, at variance with the previous
examples, the rate function vanishes in the whole region $\rho \ge \rho_c$. 
This is due to the fact that $P(\rho)$ has a weaker dependence on $M$ with respect to
the exponential one of Eq.~(\ref{ldp}), and hence the LDP is violated for $\rho >\rho_c$.
We will comment later on that.

\subsection{Stochastic Maxwell-Lorentz particle model}
\label{maxwell}

The so-called stochastic Maxwell-Lorentz
gas~\cite{alastuey2010,gradenigo2012} consists of a probe particle
of mass $m$ whose velocity $v$ changes due to the collisions with bath
particles, of mass $M$ at temperature $T$, and due to the acceleration
produced by an external force field ${\cal E}$. Collisions with the scatterers
change instantaneously the probe's velocity from $v$ to $v'$ and we
assume the simple collision rule
$v'=V$, 
where $V$ is the velocity of the scatterer, drawn from
a Gaussian distribution:
\begin{equation}
P_{\scatt}(V)= \sqrt{\frac{M}{2\pi T}} e^{-\frac{MV^2}{2T}}.
\label{eq:gaussian}
\end{equation}
The scatterers play the role of a thermal bath in
contact with the probe particle.  This model is a particular case of a
more general class of systems studied
in~\cite{alastuey2010,gradenigo2012}.  During a time $\tau$ between
two consecutive collisions, the probe performs a deterministic
motion under the action of the field $\mE$.  We assume that the
duration of flight times $\tau$ is exponentially distributed
$P_\tau(\tau)= \frac{1}{\tau_c} \exp(-\tau/\tau_c)$ and independent of
the relative velocity of the particles.
The system reaches a non-equilibrium stationary state characterized by a total entropy production $\Delta s_{\tot}$,
associated with the velocity $v(t)$, defined as
\begin{equation} 
\Delta s_{\tot}(t) = \ln \frac{\prob(\{v(s)\}_0^t)}{\prob(\overline{\{v(s)\}_0^t})},
\label{epdef}
\end{equation}
where $\prob(\{v(s)\}_0^t)$ and $\prob(\overline{\{v(s)\}_0^t})$ are,
respectively, the pdf of a path $\{v(s)\}_0^t$ spanning the time
interval $[0,t]$ and of the time-reversed path
$\overline{\{v(s)\}_0^t}=\{-v(t-s)\}_0^t$~\cite{lebowitz1999}. 
This fluctuating quantity takes contributions at any time and is therefore extensive in $t$. 
In this example it plays the role of the collective variable $N$, and $t$ plays the
role of the number $M$ of elements contributing to it.  

\begin{figure}[ht]
\centering
\rotatebox{0}{\resizebox{.45\textwidth}{!}{\includegraphics{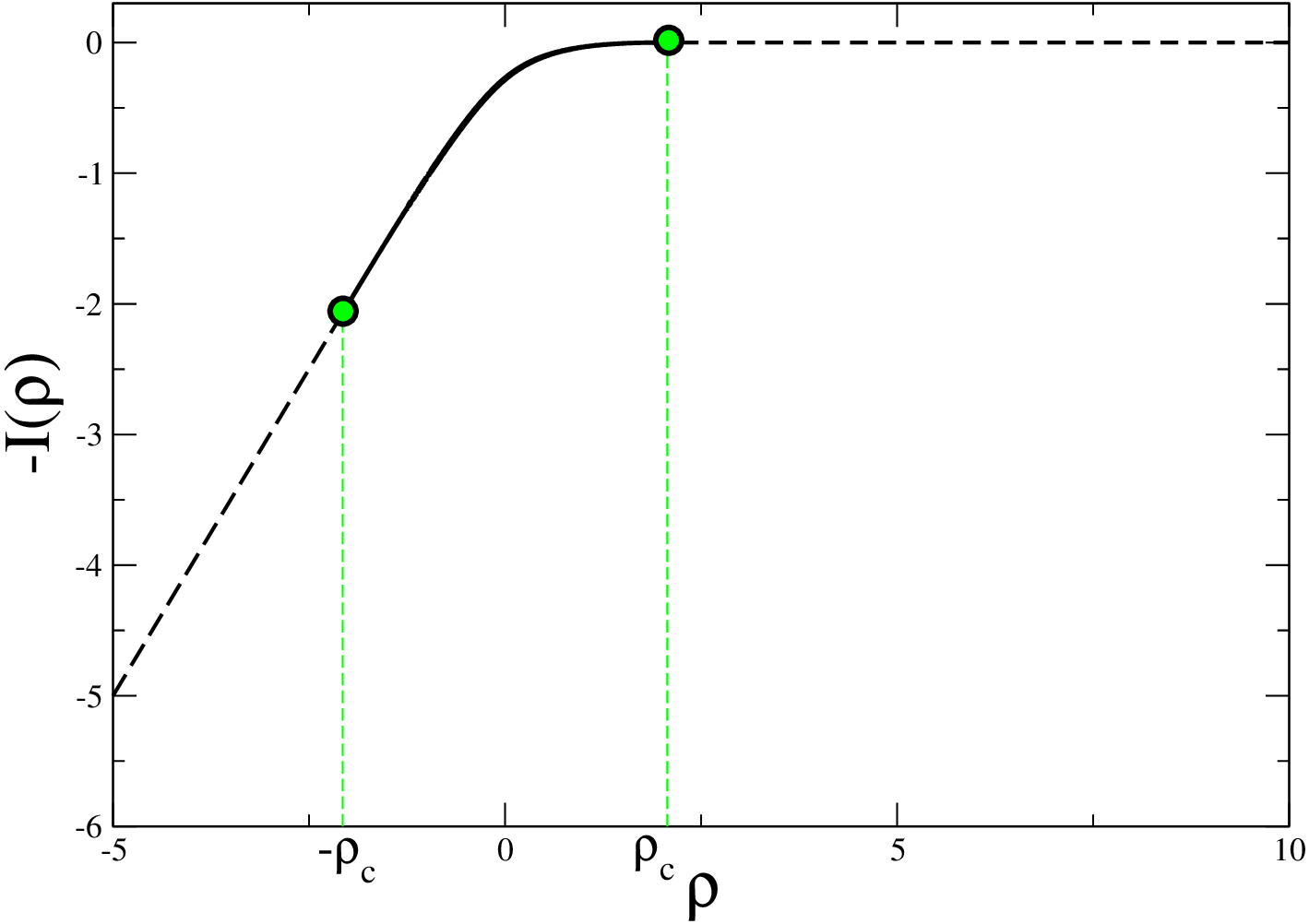}}}
\caption{The rate function $I(\rho)$ of the quantity $\rho=\Delta
  s_{\tot}/t$ for the Maxwell-Lorentz gas model~\cite{gradenigo2013},
  computed analytically in the
  limit $t\to\infty$.}
\label{fig_touch}
\end{figure}

The rate function $I(\rho)$ of the quantity $\rho=\Delta s_{\tot}/t$
was studied in~\cite{gradenigo2013} by means of numerical simulations
for finite times and analytically in the limit $t\to\infty$. This
quantity is shown in Fig.~\ref{fig_touch}, where
$\rho_c=m\tau_c\mE^2/\theta$, with $\theta=T m/M$ playing the role of
an effective temperature~\cite{PSV2017}. 
Also in this case, as for the urn model, $I(\rho)$ vanishes and $P(\Delta s_{tot})$
does not satisfy a standard LDP for $\rho >\rho_c$. Indeed it can be shown that
 the far positive tail of $P(\Delta s_{tot})$ scales
exponentially with $\sqrt{t}$ rather than with
$t$~\cite{gradenigo2013}, how it should be if the LDP (Eq.~(\ref{ldp}))
holds. Recently, the nature of the singularities in
$I(\rho)$ and their physical meaning have been
thoroughly discussed in a similar model in~\cite{gradenigo2018}, where the observed
non-analytical behaviors have been related to a first-order dynamical
phase transition.

\subsection{Some other models}
\label{kink}

We have discussed above some models where a singular probability
distribution was found.  
All these cases can be grouped into two classes: the first contains the cases
where the rate function is well defined, although it contains some non-analyticity point.
The examples of Secs.~\ref{thegaussianmodel},\ref{heatlargen} behave in this way.
The second class is the one represented by the urn model, where the probability distribution is
still singular, but the the rate function is not defined in a certain region (that is to say it vanishes identically).
The Maxwell-Lorentz gas is an example where the two behaviors are exhibited in different regions
of the fluctuation spectrum.
 
Beyond the cases discussed before, other examples of singular behavior include the probability
distribution of the work done by active particles~\cite{cagnetta2017},
of the heat exchanged by harmonic oscillators during a quench with a
thermal bath~\cite{crisanti}, of the magnetization in the spherical
model~\cite{magnetic1,magnetic2}, of the displacement of a Brownian walker with
memory~\cite{memory}, of the work done in a quantum
quench~\cite{qquench}, and many others~\cite{touchette2009,zan,oth1,oth2,oth3,oth4,oth5,oth6,oth7,oth8}.

We also mention the case where the singularity appears as a ``kink''
in zero in the probability distribution, showing a linear regime for
negative values. This behavior has been observed in the distribution
of the entropy production and of other currents for a driven particle
in periodic potentials~\cite{mehl,speck,nyawo}, in a molecular motor
model, described in~\cite{lacoste}, and in the experimental results
reported in~\cite{kumar}, where the large deviation function of the
velocity of a granular rod was measured. In general, the presence of
the kink can be related to different physical
mechanisms~\cite{fischer}, such as intermittency~\cite{budini},
detailed fluctuation theorem~\cite{dorosz}, and dynamical phase
transitions~\cite{garrahan}.

\section{General features of singular probability distributions}
\label{general}

In this section we will discuss some general properties of singular probability distributions
observed in the different models mentioned above,
focusing on the common physical interpretation and on the underlying mathematical  structure.

\subsection{Duality} \label{duality}

The singular behavior of the probability distribution seen in the examples of the previous section
has an interpretation akin to the occurrence of phase transitions in ordinary critical phenomena.
In order to discuss this point we can refer to the Gaussian model as a paradigm. 
The partition function is
\be
Z=\int \delta \varphi \,{\cal P}(\varphi),
\label{part}
\ee
where ${\cal P}$ is the probability of microscopic configurations as specified by the field ${\varphi}$.
For instance, in a canonical setting it is ${\cal P}(\varphi)=Z^{-1}\exp[-\beta {\cal H}(\varphi)]$, where $\beta$ is 
the inverse temperature $\beta =1/(k_BT)$; in this case $Z$ depends on $T$ and $V$, the volume.
On the other hand the probability of the collective variable $N$ of Eq.~(\ref{defvar}) can be written as
\be
P(N)=\int \delta \varphi \,{\cal P}(\varphi) \, \delta \left (\int _V d\vec  x \varphi ^2(\vec x)-N\right ).
\label{probpart}
\ee In view of Eq. (\ref{part}), one can recognize
Eq. (\ref{probpart}) as a partition function as well.  However this is
not the partition function of the original model that is, in this
example, the Gaussian one. Instead, $P(N)$ in Eq. (\ref{probpart}) can
be interpreted as the partition function of a {\it dual} system that
can be obtained from the original one upon removing all the
configurations such that the argument of the delta function in
Eq. (\ref{probpart}) does not vanish.  In other words, this is the
model one arrives at upon constraining configurations in a certain
way. In this case the requirement is that the variance of $\varphi$
must equal a given value $N$. Such a system, a Gaussian model with a
constraint on the variance, is the spherical model of Berlin and Kac
\cite{berlinkac}. 

The equilibrium properties of the spherical model are exactly
known. For fixed $N$, there is a phase-transition at a critical
temperature $T_c$, from a disordered phase for $T\ge T_c$ to an
ordered one below $T_c$. Equivalently, still in the Berlin-Kac model, 
if one keeps $T$ fixed, the
transition occurs changing the variance $N[\varphi]$ defined in
Eq. (\ref{defvar}) upon crossing a critical value $N_c$. The ordered
phase is found for $N>N_c$, in this case. The presence of such a phase
transition crossing $N_c$ determines a singularity of the partition
function $P(N)$ of the spherical model (Eq. (\ref{probpart})) at
$N=N_c$. However the same quantity $P(N)$ is also the probability
distribution of the quantity $N[\varphi]$ in the context originally
considered, the Gaussian model. This explains what one observes in
Fig. \ref{fig_gauss}.  $N_c$ is the value of $N$ marked by a dot in
this figure, where the singular behavior shows up.

This dual interpretation of $P(N)$, as a probability distribution of a
collective variable in the original model, or as a partition function
in a dual model, may help to understand why singularities are manifested in
the probability distributions. Indeed, if one asks the question: {\it why a simple model without phase
transition, such as the Gaussian model, exhibits a non trivial
singularity in the probability distribution $P(N)$?}, the answer can be
that, although the original model is quite simple, the dual one is far
from being trivial, with a phase-transition induced by the presence of
the constraint. This originates an anomalous behavior in the fluctuation
spectrum of the original model.

We have discussed the fact that imposing a constraint to the Gaussian
model we change the system into a dual one that is interested by a phase transition, 
since this is the spherical model. Is this an isolated example or this feature has
some generality? The answer is that it occurs quite often.  Besides
the above mentioned spherical model, another well known example where
the same mechanism is at work is the perfect boson gas.  There is no phase transition
in a gas with a non conserved number $N$ of bosons, as in the case of
photons, but if the number $N$ of particles is fixed Bose-Einstein
condensation happens. The partition function of the conserved bosons,
for a given volume and temperature, has a singularity at a certain
value of the boson number $N=N_c$ (or density). This singularity
corresponds to the critical number of particles below which the
condensed phase develops. According to the duality principle discussed
above, this implies that the probability distribution of the number of
bosons in a system of, say, photons, where this number is allowed to
fluctuate, will be singular at the same value $N_c$ of the random
variable $N$~\cite{zan}. The very urn model is another instructive example.
One can consider a model, dual to the one discussed in Sec.~\ref{urnmodel}, where the
total number of balls is conserved~\cite{noldp8}. Marbles can only be exchanged among boxes 
and their density $\rho$ is an external control parameter. This model is known to
be interested, for $k>2$, by a phase transition crossing $\rho =\rho_c$. Notice that,
since $\rho $ is a control parameter, having $\rho >\rho_c$ in this dual model is not
a rare event (as in the model introduced in Sec.~\ref{urnmodel}). A similar situation is
found in related models such as the zero range process~\cite{noldp8,noldp5,oth4}.

\subsection{Condensation}

In order to see how singularities may come about in another perspective we will discuss the
phenomenon in the framework of the urn models, where the physical
meaning is probably more transparent in term of a {\it condensation}
mechanism. Something similar occurs also in the other models
considered in Sec.~\ref{examples}, regardless of the fact that the 
rate function is well defined or not.

\begin{figure}[ht]
\vspace{2cm}
\centering
\rotatebox{0}{\resizebox{.45\textwidth}{!}{\includegraphics{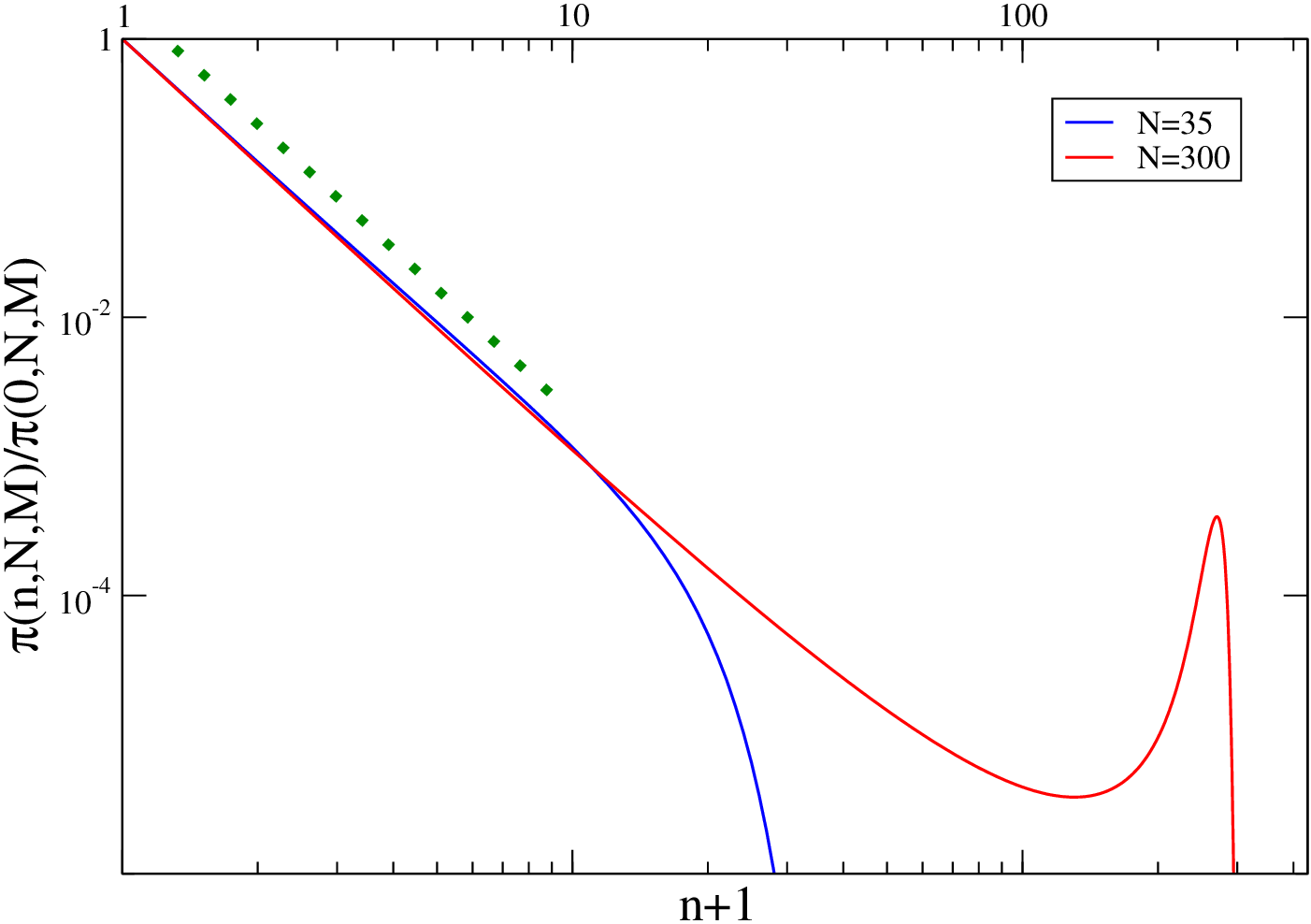}}}
\caption{The function $\pi(n,N,M)$ is plotted, for $k=3$ and $M=100$, against $n+1$ for two values of $N$:
$N=35$, corresponding to a case without condensations, and $N=300$, corresponding to a 
condensed situation. The dotted green curve is the power-law $x^{-k}$.}  
\label{fig_pi}
\end{figure}

Let us consider the conditional probability $\pi (n,N,M)$ that one of
the $M$ {\it a priori} equivalent urns contains $n$ particles, given
that there are $N$ particles in the whole system.  This quantity can
be evaluated exactly and is shown in Fig. \ref{fig_pi} (normalized by
its value in $n=0$ to better compare curves with different $N$ in a
single figure). Let us discuss its properties.  First of all $\pi$
vanishes for $n>N$, since it is impossible that an urn contains more
particles than the whole system. Secondly, for small $n$ one has $\pi
(n,N,M)\propto p(n)$ (dotted green line in Fig. \ref{fig_pi}).  This means
that, as far as very few particles are stored in the tagged urn, the
condition on the total number of balls is irrelevant.

More interestingly, at large $n$, $n\lesssim N$, different behaviors
are observed in the region of relatively small $N$, and in the one
with relatively large $N$, exemplified by $N=35$ and $N=300$,
respectively, in Fig.~\ref{fig_pi}. In the former case $\pi$ is exponentially
damped at large $n$, meaning that accommodating many particles in a
single urn is probabilistically very unfavorable.  In the latter case
there is a peak at a value of $n$ or order $N$. This means that a
significant fraction of the total number of particles is located in a
single urn. This is the condensation phenomenon.  (We will see in the
next section that in this particular model it occurs when $k>2$ and
for sufficiently large densities).  The essence of a condensation
phenomenon is that a given quantity is not fairly distributed among
many degrees of freedom, but is concentrated in a single one. This is
particularly clear in the urn model, where one particular urn
contains a macroscopic fraction of balls.

One easily realizes that something similar occurs in the other example
models discussed in Sec.~\ref{examples}. For instance, in the model of
Sec.~\ref{thegaussianmodel}, writing $N[\varphi]$ in terms of the
Fourier components $\varphi _{\vec k}$ of the field $\varphi$ as \be
N[\varphi]=\frac{1}{V}\sum _{\vec k}\vert \varphi _{\vec k} \vert ^2,
\label{vark}
\ee one can show that while for $N\le N_c$ (or equivalently $\rho \le
\rho _c$) all the Fourier components add up to realize the sum in
Eq.~(\ref{vark}) in a comparable way, for $N>N_c$ the term with $k=0$
alone provides the most important contribution to the sum. A similar
mechanism, with the dominance of the $k=0$ term, is also at work in
the example of Sec.~\ref{heatlargen}.  In the Maxwell-Lorentz particle
model (Sec.~\ref{maxwell}) one has that normal entropy fluctuations
are formed by the addition of many contributions associated to many
short flights of the probe particle.  However above the critical
threshold $\rho _c$ they are associated to a single event which is
responsible for a macroscopic contribution to the entropy
production. This event is a long flight of the probe particle with no
collisions with the scatterers.  For more details and a very accurate
analytical description of these kinds of behaviors in a similar system
re-framed in the context of active particles, see the recent
work~\cite{gradenigo2018}.

\subsection{Mathematical mechanism}

In the previous section we have discussed the phenomenon of condensation on physical grounds.
In this section we show the mathematical mechanism behind. We will give a description as simple as
possible, without presumption of mathematical rigor, in the framework of the urn model.  

\begin{figure}[ht]
    \centering
	\vspace{1cm}
   \rotatebox{0}{\resizebox{.45\textwidth}{!}{\includegraphics{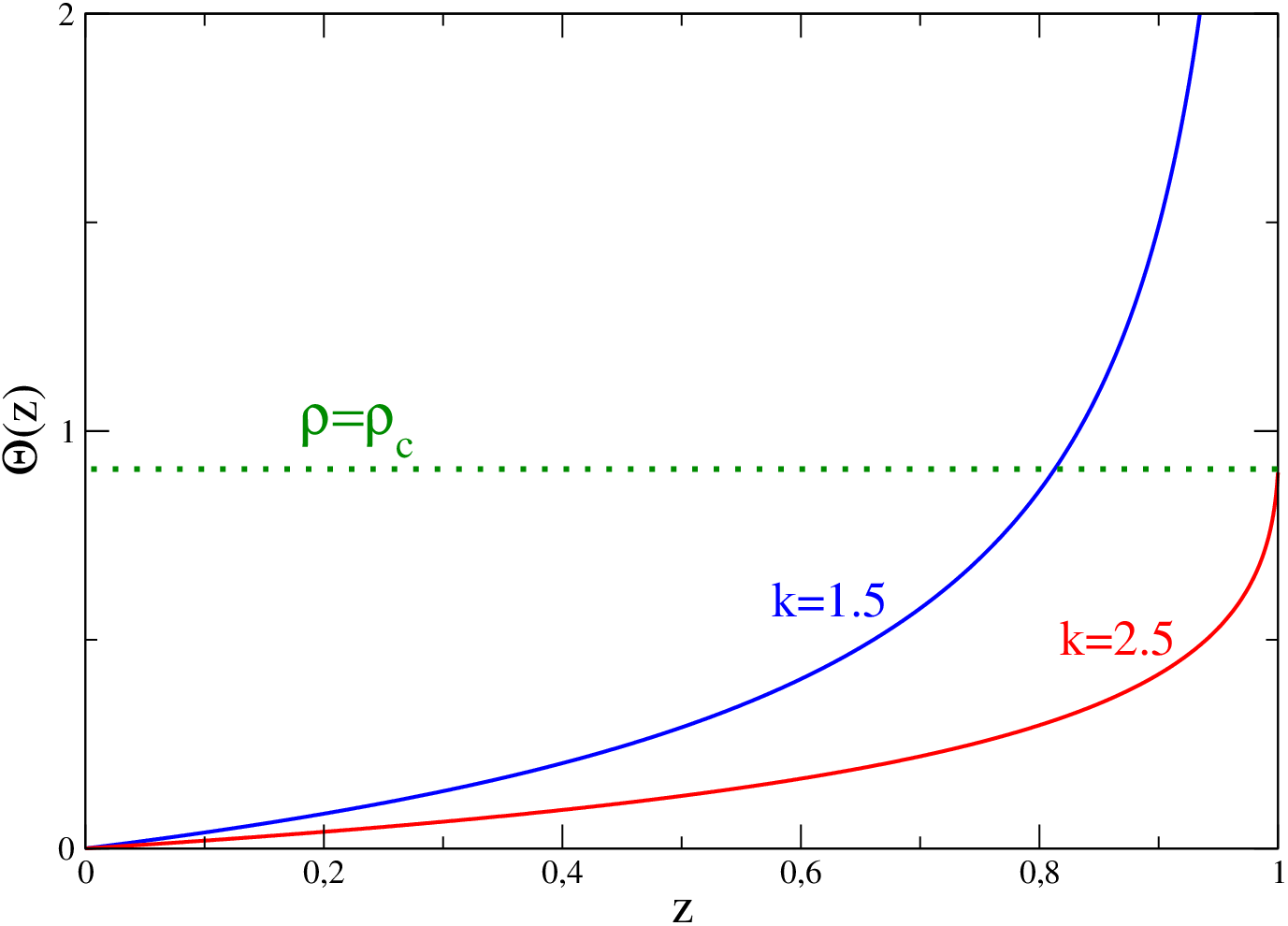}}}
    \caption{The function $\Theta(z)$ is shown for $k=1.5$ and $k=2.5$.}
\label{selfurne}
\end{figure}

The probability distribution of the total number of particles $N$ reads
\be
P(N,M)=\sum _{n_1,n_2,\dots,n_M}p(n_1)p(n_2)\dots p(n_M)\,\delta_{\sum _{m=1}^Mn_m,N}\quad ,
\label{proburne}
\ee
where $\delta_{a,b}$ is the Kronecker function and in the leftmost sum 
the variables $n_1,n_2,\dots,n_M$ run from $0$ to $\infty$.
Using the representation
\be
\delta_{a,b}=\frac{1}{2\pi i}\oint dz \, z^{-(b-a+1)}
\ee
of the $\delta $ function one arrives at
\be
P(N,M)=\frac{1}{2\pi i}\oint dz \, e^{M[\ln Q(z)-\rho \ln z]},
\label{eqoint}
\ee
where 
\be
Q(z)=\sum _{n=0}^\infty p(n)z^n,
\ee
and we have confused $\frac{N+1}{M}$ with $\rho=N/M$ for large $M$.
Still for large $M$, the integral in Eq. (\ref{eqoint}) can be evaluated by the steepest descent method as
\be
P(N,M)\simeq e^{-MR(\rho)},
\ee
where 
\be
R(\rho)=-\ln Q[z^*(\rho)]+\rho \ln z^*(\rho),
\ee
with $z^*$ the value of $z$ for which the argument in the exponential of Eq. (\ref{eqoint}) has its maximum value.
This in turn is given by the following implicit saddle-point equation
\be
\Theta (z^*)=\rho,
\label{saddle}
\ee
where 
\be
\Theta (z^*)=z^* \frac{Q'(z^*)}{Q(z^*)}.
\ee
Let us study this equation. Clearly, it must be $z\le 1$ in order for the sums hidden in $Q$ and $Q'$ to converge.
It can also be easily seen that $\Theta(0)=0$ and that this function increases with $z$ up to
\be
\Theta(1)=\left \{ \begin{array}{ll}
\infty, & k\le 2 \\
\Theta_M, & k>2, \end{array}
\right .
\ee
where $\Theta_M$ is a finite positive number. The function $\Theta(z)$ is shown in Fig. \ref{selfurne}, for two values
of the parameter $k$. As it is clear from this figure, for $k>2$ the saddle point equation (\ref{saddle})
admits a solution only for $0\le \rho \le \rho_c=\Theta (1)$. It is trivial to show that 
$\rho_c\equiv \langle \rho \rangle=\sum _n np(n)$. 
However nothing prevents fluctuations with
$\rho >\langle \rho\rangle$ to occur.
How can we recover the model solution for $\rho >\langle \rho \rangle$?
We know that for such high densities urns are no longer equivalent: there is one -- say the first --
which hosts an extensive
number of particles and condensation occurs. In a physically oriented approach, we  can 
take into account this fact by writing, in place of
Eq. (\ref{proburne}), the following
\be
P(N,M)=M\sum _{n_1=0}^\infty p(n_1)\sum _{n_2,n_3,\dots,n_M}p(n_2)p(n_3)\dots p(n_M)
\,\delta_{\sum _{m=2}^Mn_m,N-n_1}\quad .
\label{proburnecond}
\ee
The factor $M$ in front of the r.h.s. stems from the fact that there are
$M$ ways to chose the urn (denoted as 1) to be singled out.
Repeating the mathematical manipulations as in Eqs. (\ref{proburne},\ref{eqoint}),
but only on the sum $\sum _{n_2, n_3, \dots,n_M}\dots$, one arrives at
\be
P(N,M)=\frac{M}{2\pi i}\sum _{n_1=0}^\infty p(n_1)\oint dz \, e^{M[\ln Q(z)-(\rho -\frac{n_1}{M})\ln z]}.
\label{eqoint1}
\ee
Evaluating the integral with the steepest descent method, the saddle point equation is now
\be
\Theta (z^*)=\rho -\frac{n_1}{M}.
\label{saddlecond}
\ee
Notice that in a {\it normal} situation, where condensation does not occur, in the thermodynamic limit 
where $M\to \infty$ with fixed $\rho$, the typical number of particles 
in a single urn does not depend  on the number of urns. Therefore the last term on the r.h.s. of Eq. (\ref{saddlecond})
is negligible and one goes back to the previous saddle point equation (\ref{saddle}).
However, when condensation occurs (i.e. with $k>2$ and $\rho >\langle \rho \rangle$) the only possibility to close the model
equations is to have the last term in Eq. (\ref{saddlecond}) finite. In conclusion one has
\be
\left \{ \begin{array}{llr}
z^*<1, & \frac{n_1}{M}\simeq 0 & \mbox {no condensation} \\
z^*=1, & \frac{n_1}{M}=\rho -\langle \rho \rangle & \mbox {condensation}.
\label{zstarwhole}
\end{array} \right .
\ee

Clearly we are in the presence of a phase-transition resembling the ferro-paramagnetic or the gas-liquid transitions. 
There are two phases with qualitatively different behaviors. However, at variance with usual phase transitions,
here the parameter producing the transition is not an external one that can be varied at will,
but the value of the spontaneously fluctuating variable $N$. 
Another difference with usual phase transitions is the fact that here there is no interaction among urns. 
Despite that, urns are not completely independent due to the constraint over the number of particles represented
by the Kronecker function in Eqs. (\ref{proburne},\ref{proburnecond}). This constraint can be regarded as an effective interaction determining the transition (it can be easily seen, in fact, that without such conservation there is no transition).

Notice that it is $n_1/M   = 0$ in the normal phase
and $n_1/M \neq 0$ in the condensed phase, therefore this quantity represents the order parameter of the transition.
Despite the fact that {\it a priori} the system (i.e. the Hamiltonian) is invariant under a permutation of the urns, namely all boxes are equal, this property is not shared by the physical realization of the actual state of the system when condensation occurs, since one urn behaves very differently from the others. We are in the presence of  
spontaneous symmetry breaking. 

As a final remark, let us note that the phenomenon of condensation in the sum 
of many identically distributed variables is not specific to an algebraic decay of $p(n)$, or to the
discrete value of the variable $n$. Indeed it is found~\cite{oth7} that it occurs provided that 
$\sum _n n p(n)<\infty$. Condensation in the presence of a stretched exponential $p(n)$, for instance,
has been discussed in~\cite{nagaev1,nagaev2}. Finally, we mention the fact that in the context of L\'evy flights
the phenomenon of condensation is usually referred to as the {\it big jump principle}~\cite{chistyakov}.

\subsection{Fluctuation Relation}

The Fluctuation Relation is one of the few general results of
non-equilibrium statistical mechanics, expressing an asymmetry property
of the fluctuations of some extensive (in time or in number of degrees of
freedom) quantities $N$ \cite{umberto}.  The FR
reads
\be 
\frac{P(N/M=\rho)}{P(N/M=-\rho)}=e^{cM\rho+o(M)},
\label{FR}
\ee 
where
$c$ is a constant, and $o(M)$
stands for sub-linear corrections in $M$. 
Usually, the exponential form of the FR is related to two properties
of $P(N/M)$: (i) it satisfies a LDP (\ref{ldp}), and (ii) the rate
function $I(\rho)$ has the symmetry: \be I(-\rho)-I(\rho)=c\rho.
\label{eqsym1}
\ee These two conditions, with $I(\rho)$ different from $0$ and
$\infty$, are known to be sufficient for $N/M$ to satisfy a FR (see,
e.g., \cite{touchette2009} and references therein).

It is interesting to consider the validity of an FR in the case of
probability distributions with singularities.
First, let us note that, when the singularity appears
in zero, as in the case of the ``kink'' mentioned in Sec.~\ref{kink},
then the validity of an FR is clearly not affected by the singularity.
More in general, the FR can also be satisfied by a pdf for which a standard
(namely, with a leading exponential scaling in $M$) LDP does not
hold. This can be observed for instance in the driven Maxwell-Lorentz
gas described in section \ref{maxwell}. In this model it has been
shown \cite{gradenigo2013} that the entropy production calculated over
a time $t$ satisfies an FR, even though the far positive tail of its
pdf scales exponentially with $\sqrt{t}$ rather than $t$. In this case
the validity of the FR can be exploited to extract some information on
the behavior of the probability distribution in the regions where the
stretched-exponential scaling takes place.

The FR~(\ref{eqsym1}) in the presence of a singular rate function has
been also observed~\cite{heatexch}, besides the already mentioned Maxwell-Lorentz case , in some large time limit for the
exchanged heat, in the large-${\cal N}$ model of
Sec.~\ref{heatlargen}.  More recently, it has been
shown~\cite{crisanti} that the rate function of the heat exchanged by
a set of uncoupled Brownian oscillators with the thermostat during a
non-stationary relaxation process does not satisfy an FR in the
form~(\ref{FR}).  Although, even in this case, the rate function shows
a singular behavior in the limit of a large number of degrees of
freedom, the lack of a standard FR is not necessarily related to the presence of
the singularity.


\section{Some peculiarities of singular distributions}
\label{peculiar}

\subsection{Giant response}

Generally, the behavior of a collective quantity such as the empirical mean (\ref{defN})
is not substantially altered if, for large $M$, the properties of only one out of $M$ variables is
slightly modified. For instance one does not expect to observe any significant change in the 
thermodynamic properties of a gas of identical molecules if one is replaced with another of
a different substance.  This is because the collective properties are determined by the synergic 
contribution of a huge number $M$ of constituents, and hence the features of a single molecule are 
negligible. This is true not only for the typical properties but also for the fluctuation distribution. 
However, the situation can be dramatically different when singular probability distributions 
enter the game. 

Let us show this with the prototypical example of the urn model.  We
consider a slightly modified version of the model defined in section
\ref{urnmodel}, where a single variable, say $n_\ell$, is distributed
as in Eq. (\ref{urnmicroprob}) but with an exponent $k_\ell$ that may
be different from the one, $k$, of all the remaining ones.  Let
us now look at Eqs. (\ref{eqoint1}, \ref{saddlecond},
\ref{zstarwhole}). In a situation where condensation does not occur,
as we remarked earlier, the effect of a single variable is negligible,
the first line of Eq. (\ref{zstarwhole}) applies and hence
$\frac{n_m}{M}\simeq 0$, for any $m$.  On the other hand, in the
presence of condensation, the second line of Eq. (\ref{zstarwhole})
holds.  In the case of equally distributed variables condensation
occurs with equal probability in any of the urns. However, if the
$\ell$-th variable behaves differently, one has to understand if the
condensing variable could be the $\ell$-th, or any of the remaining
ones. Both the cases can occur, depending on the values of the
exponents $k$ and $k_\ell$.

For $k_\ell>k>2$ (the latter inequivalence being needed for condensation) it is $p(n_\ell=n)\ll p(n_m=n)$
for large $n$ (with $\ell \neq m$). Hence the condensation phenomenon, which occurs by letting a huge
amount of particles occupy a single urn, is unfavoured in the $\ell$-th urn. 
The situation in this case is analogous to the one discussed before with equally distributed variables,
i.e. with $k_\ell = k$. However for $k>k_\ell>2$ the opposite occurs, the condensing variable is the $\ell$-th.
Hence Eq. (\ref{eqoint1}) applies with $n_1$ replaced by $n_\ell$.
One sees from Eq. (\ref{eqoint1}) that, when condensation occurs, $P(N,M)$ is proportional to   
$p(n_\ell)$. Since $k_\ell \neq k$, $P(N,M)$ turns out to be different from the one found for
equally distributed variables. Hence, in this case, an even small change of the properties of a single variable can
trigger the form of the probability distribution of the collective variable $N$, a fact that is sometimes referred to
as {\it giant response}. 

\begin{figure}[ht]
    \centering
	\vspace{1cm}
   \rotatebox{0}{\resizebox{.45\textwidth}{!}{\includegraphics{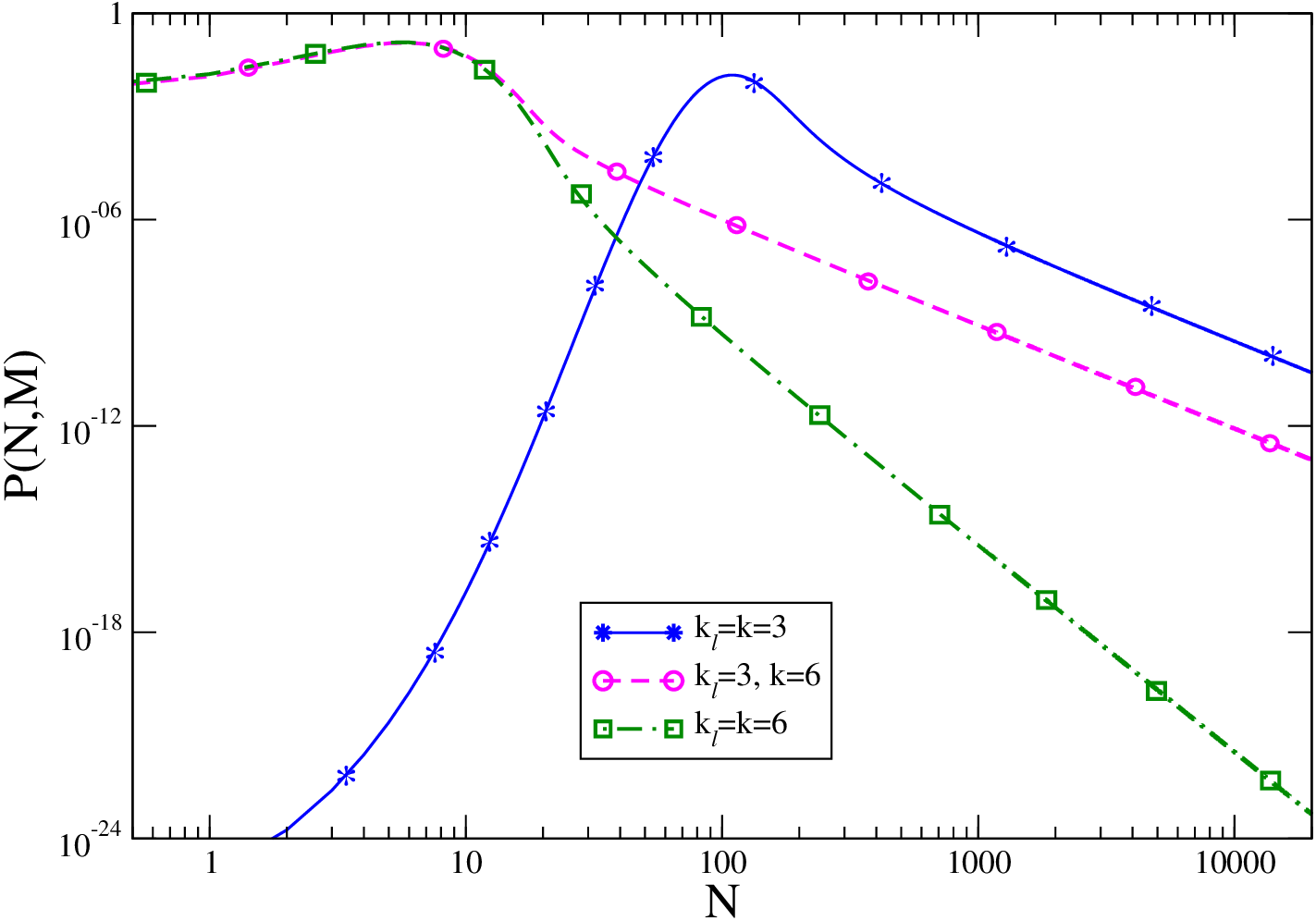}}}
    \caption{$P$ is plotted for $M = 333$ and the three different choices (see text) i) $k_\ell \equiv k = 3$, 
    continuous blue with asterisks, ii) $k_\ell \equiv k = 6$, dot-dashed green with squares and iii) $k_\ell =3$ ,  $k=6$,
    dashed magenta with a circles.}
\label{fig_giant_resp}
\end{figure}

This is illustrated in Fig. \ref{fig_giant_resp}.
Here $P(N,M)$ is compared for three different choices of the exponents $k,k_\ell$.
The continuous blue curve with asterisks refers to the case i) with identically distributed variables with $k_\ell=k=3$.
Similarly, the dot-dashed green curve with squares corresponds to the situation with ii) $k_\ell=k=6$.
Instead, the dashed-magenta curve with circles corresponds to non-identically distributed variables with iii) $k_\ell=3$ and $k=6$.
Notice that in the region to the left of the maximum, where condensation does not occur 
(because in this region $\rho < \langle \rho \rangle$), the curves of the cases ii) and iii) coincide.
This nicely shows that in the absence of condensation the shift of the properties of a single variable does not
influence the collective behavior of the system. 
For $\rho > \langle \rho \rangle$, on the other hand, the form of $P$ drastically changes in going from ii) to iii),
namely by perturbing the properties of one single variable. Even more impressive, the slope of the curve
for case iii) is the same as that of case i), showing that this feature is dictated by the sole properties of the
variable, $n_\ell$, which in case iii) behaves as in i).

\subsection{Development of a singular fluctuation}

We have seen in Sec. \ref{duality} that a singularity in the
probability distribution can be interpreted as a phase transition
occurring at a {\it critical} value of $\rho$, playing the role of a
control parameter.  The analogy can be pushed a step further. When a
system is prepared in a certain equilibrium state and then a control
parameter is changed as to make it cross a phase transition, the
ensuing dynamics can be slow and characterized by a dynamical scaling
symmetry associated most of the times with an ever growing length
scale \cite{bray94,bray942,bray943,bray944}. Typical examples are magnets and binary systems
quenched across the critical temperature, and glassy systems.

Building on the analogy above, one might expect something similar to
happen if one prepares a system with a singular $P(N)$ in a state such
that the fluctuating collective variable $N$ takes a definite value
$N_0$ on one side (say the left) of the critical value $N_c$ where the
singularity takes place. If the system is then left to evolve freely,
all possible fluctuations will take place, including those on the
other side (say the right) of the singularity. Due to the duality
principle, this process should occur in a way akin to the kinetics of
a system brought across a phase-transition. Hence slow evolution and
dynamical scaling should be observed. This has actually been shown to
be the case, as we discuss below.

\begin{figure}[ht]
\vspace{2cm}
\centering
\rotatebox{0}{\resizebox{.45\textwidth}{!}{\includegraphics{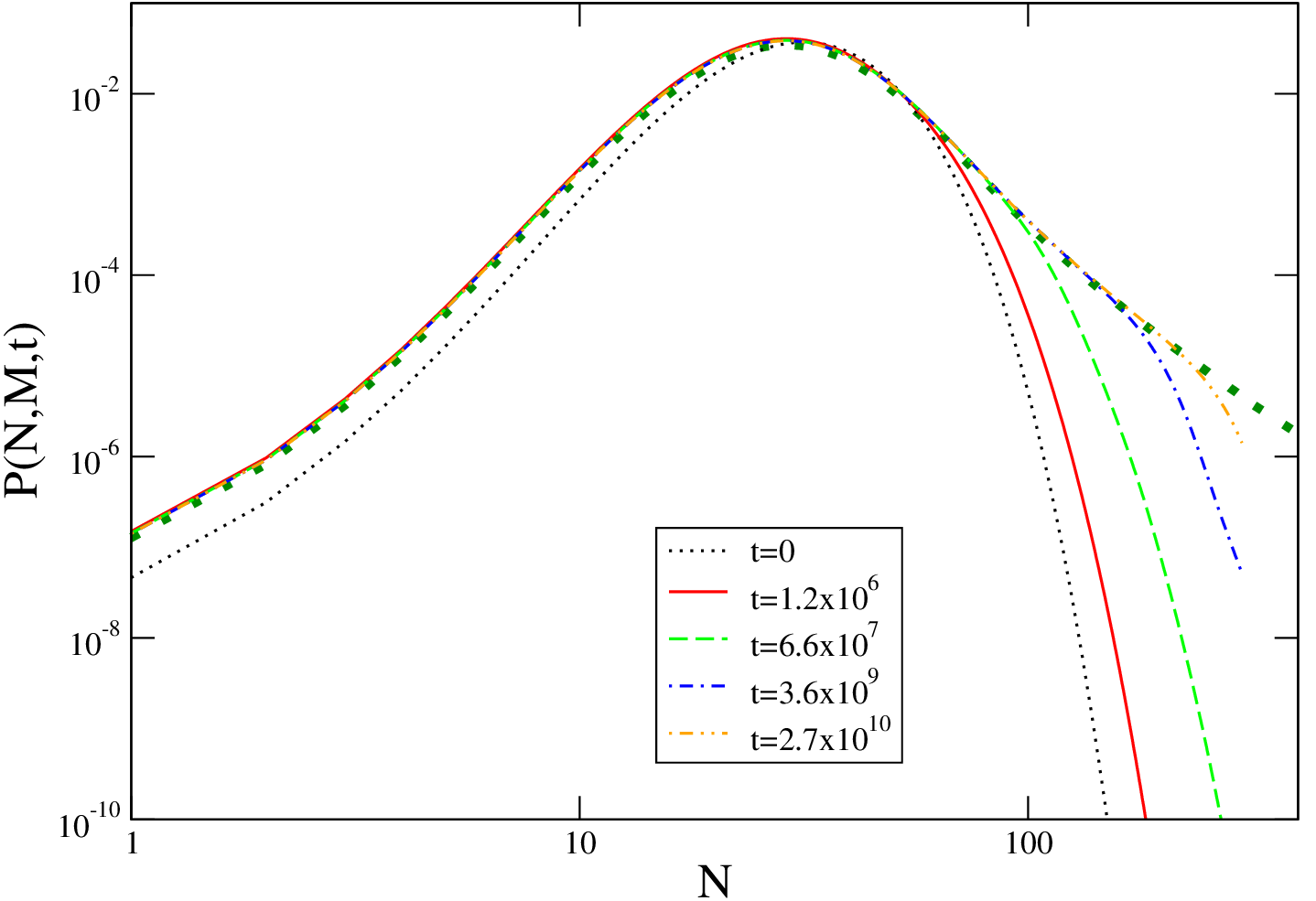}}}
\caption{The probability $P(N,M,t)$ with $k=3$ is plotted against $N$ with double logarithmic scales for different times
(see key), exponentially spaced. The dotted green line is the asymptotic form.}  
\label{fig_dyn_urns}
\end{figure}

Upon supplementing the urn model of Sec. \ref{urnmodel} with a kinetic rule allowing the system to exchange
single particles with an external reservoir in such a way that the stationary
occupation probability of any urn is given
by Eq. (\ref{urnmicroprob}), one can solve exactly \cite{corb_dyn} the evolution of a system whose initial state
is such that condensation is not present. In the following we will discuss the case 
in which the initial value of the density is  $\rho =\langle \rho \rangle$.
Starting from this configuration, corresponding to an initial form $P(N,M,t=0)$ of the probability distribution
of the collective variable, the system will evolve as to produce all the allowed fluctuations. Hence 
$P(N,M,t)$ becomes time-dependent. 
Clearly, for long times it is expected to approach the stationary value 
$P(N)$, with the singular behavior already discussed.  This curve is plotted in Fig. \ref{fig_dyn_urns},
with a dotted green line. 

In this figure one sees that the time evolution of the probability $P(N,M,t)$ towards this asymptotic form is 
much different on the two sides of the singularity. For $N< \langle N\rangle$, in the {\it normal } region without
condensation, the evolution is fast and the asymptotic form of the probability is attained at relatively short times. Indeed, already the red curve for $t=1.2$x$10^6$ is indistinguishable from the stationary form and
increasing time does not change anything.
Conversely, the evolution is slow in the condensing region for $N>N_c$. Here one sees that, at any time, the 
asymptotic form is only attained up to a value $N=\nu (t)$, beyond which $P(N,M,T)$ drops much faster than 
what expected asymptotically. It can be shown that $\nu (t)$ grows indefinitely in an algebraic way, much in the
same way as a characteristic growing length does in systems quenched across a phase transition.
In addition, a dynamical scaling symmetry can be shown to be at work also in this case.
The origin of this slow kinetics is clearly due to the difficulty to condense a huge amount of particles in 
a single urn by exchanging single particles with the reservoir.
 
\subsection{Observability}

In the previous sections we discussed some peculiar properties of singular distribution functions.
A natural question is if such features can be observed in practical situations.
Indeed, the non-analycities of the probability distributions are observed in the regime of {\it large deviations},
namely outside the range of small fluctuations which are generally described by the central limit theorem and 
are more likely to be observed. 

To make more clear this point let us make reference to the Gaussian
model and, specifically, to Fig. \ref{fig_gauss}. In this case, in
order to detect singular deviations, $\rho =\rho
_c$ must be exceeded. Namely, the system has to move quite far from the most likely observed
value -- the maximum of the distribution. If the LDP (\ref{ldp}) holds
(it does so in this model) the possibility to observe such a large
fluctuation is extremely small already for moderately large values of
the number of constituents $M$ (or volume $V$), due to the exponential damping 
in $M$ expressed by Eq.~(\ref{ldp}). But the situation is
different if the LDP is violated. This occurs, for instance, in the
urn model or in the Maxwell-Lorentz gas, in the fluctuation range where the rate function vanishes. 
In the former model one can easily check from Eqs. (\ref{eqoint1},
\ref{saddlecond}, \ref{zstarwhole}) that the LDP is obeyed in the
non-condensing regime but it is violated when condensation
occurs. In fact, it is trivial to see  that with $z^*=1$ the saddle
point evaluation of the integral in Eq. (\ref{eqoint1}) gives an
exponential with an argument that is identically vanishing.  As a
consequence fluctuations away from the average are no longer damped
exponentially in $M$, but only as $M^{1-k}$ (keeping $\rho $
fixed). This is why the rate function of the model vanishes in the
whole sector $\rho >\rho _c$ where condensation occurs (see
Fig. \ref{fig_urns}), despite the fact that $P(N,M)$ decays for $\rho
>\langle \rho \rangle$, as it can be seen in
Fig. \ref{fig_giant_resp}.  Due to this much softer decay, there is a
better chance to observe singular fluctuations in this model than in
others, e.g. the Gaussian model, where the LDP holds.  A similar
situation, with LDP violations, is observed also in the
Maxwell-Lorentz particle model (for $\rho=\Delta s_{tot}/t>\rho_c$)~\cite{gradenigo2013} and
in Bose-Einstein condensates \cite{zan}.

\section{Summary and conclusions}\label{conclusions}

In this paper we have shortly reviewed the issue of probability distributions characterized by
non-analyticities. Naively, this feature could be considered as a rare manifestation of curious
mathematical pathologies occurring in scholarly model with uncertain relations to the physical world. 
In reality, singular probability distributions have been shown analytically 
to occur in very simple and fundamental
models of statistical mechanics, such as the Gaussian one, and not only in weird
non-equilibrium states but also in equilibrium. Furthermore, they have been detected in numerical
simulations and, most importantly, also in real experiments. This widespread occurrence points towards
an underlying general mechanism for the development of singularities in the fluctuation probability.
This paper has been conceived in order to highlight and discuss, at a simple and physically oriented level,
at least some of such general features.

In the first part of the paper, after recalling basic and general concepts of probability theory we have reviewed 
some models where singular fluctuation spectra have been observed. These range from the aforementioned 
Gaussian model to the spherical limit of a ferromagnet, from the so-called urn model to a description of 
the Maxwell-Lorentz gas. In all these cases the deviations of certain collective observables are described
by non-analytical probability distributions, which, in the case when LDP holds, are characterized by the 
presence of exponential branches.  

The non-analytical behavior has been interpreted as due to the same mechanism whereby singularities
develop in the thermodynamic functions of systems experiencing phase transitions. Indeed we have discussed
the fact that a singular fluctuation distribution function can be mapped onto a thermodynamic potential of a dual 
model with a critical point. The singularity appears similarly to what one observes in thermodynamic
functions when a condensation transition is present. When such feature occurs at the level of fluctuations,
at variance with the usual examples of condensation, one speaks of condensation
of fluctuations. 

Singularities of the probability distributions can have a scarce practical relevance if they occur in regions
where fluctuations have a negligible chance to be observed. However, in some of the cases considered in
this paper the non-analytical behavior is associated to the breakdown of the large deviation principle.
As a result, large fluctuations of macrovariables have a better chance to be observed even in systems 
with a relatively large number of degrees of freedom. In this case the presence of singularities not only
can be observed, but its effects can be appreciated. Perhaps, one of the most intriguing one is the so called
giant susceptibility, whereby slightly tuning the properties of even one single component,
say a molecule of a gas, can have catastrophic consequences on the behavior of the whole system.

Non-analycity points in the probability distributions also influence the
way in which rare fluctuations are developed out of typical state where they are absent. 
Indeed, it has been shown that large fluctuations in the region where condensation occurs are
formed by means of a complex slow dynamics which resembles, once again a manifestation of a
dual behavior,  that of systems brought across a phase transition. The knowledge of the dynamical
path leading to a rare fluctuation may have important consequences in those cases when such
deviations lead to catastrophic events, as in the case of extinctions or bankruptcies.

Among the several perspectives of future studies in this context, we
mention the possibility to explore the role of correlated noise on the
large deviations, for instance in models of active particles where
some analytical results can be obtained~\cite{maggi}; the meaning of
singularities, which are related to non-equilibrium phase transitions,
within the general framework of the macroscopic fluctuation
theory~\cite{jona}; the relation between the presence of singularities
and the validity of the Fluctuation Relation for entropy production or
related quantities in more general cases; the role of
  correlations among random variables in the anomalous large
  deviations, as observed for instance in conditioned random walks~\cite{szavits}
  and Brownian motion~\cite{nickelsen}; the effect of inhomogeneous
  rates in bulk-driven exclusion processes~\cite{lazarescu}.

\vspace{6pt}

\acknowledgments{F.C. and A.S. acknowledge A. Crisanti,
  L.F. Cugliandolo, G. Gonnella, G. Gradenigo, A. Piscitelli,
  A. Puglisi, H. Touchette, A. Vulpiani and M. Zannetti for a long
  collaboration on some of the issues here
  discussed. F.C. acknowledges funding from grant PRIN 2015K7KK8L
  covering also the publication costs of the present open access
  article. A.S. acknowledges support from ``Programma VALERE'' of
  University of Campania ``L. Vanvitelli''.}

\renewcommand\bibname{References}

\end{document}